\documentclass[]{IEEEtran}
\usepackage{cite}
\usepackage{amsmath,amssymb,amsfonts,mathtools}
\usepackage{bbm, bm}
\usepackage{multirow} 
\usepackage{array}
\usepackage{graphicx}
\usepackage{hhline}
\usepackage{textcomp}
\usepackage{xcolor}
\usepackage{subcaption}
\usepackage{siunitx}
\usepackage{hyperref}

\def\BibTeX{{\rm B\kern-.05em{\sc i\kern-.025em b}\kern-.08em
    T\kern-.1667em\lower.7ex\hbox{E}\kern-.125emX}}
    
\definecolor{mehdi}{RGB}{0,0,250}

\definecolor{Samad}{RGB}{0,250,0}

\definecolor{Matti}{RGB}{250,0,0}

\begin{document}

\title{Diffusion  
Models for Wireless  
Communications
}  
 
\markboth{}
{}
\author{\IEEEauthorblockN
{Mehdi Letafati, \IEEEmembership{Graduate Student Member, IEEE,}
			Samad Ali, \IEEEmembership{Member, IEEE,}   and
		Matti Latva-aho,
		   	\IEEEmembership{Fellow, IEEE}
}
\textsuperscript{}\thanks{ 
The authors are with the Centre for Wireless Communications, University of Oulu,
Oulu, Finland (e-mails: mehdi.letafati@oulu.fi; 
 samad.ali@oulu.fi; matti.latva-aho@oulu.fi). 
}}

\IEEEaftertitletext{\vspace{-1.0\baselineskip}}

\maketitle

\begin{abstract} 
A comprehensive study on the applications of denoising diffusion models for wireless systems is provided.  
The article highlights the capabilities of diffusion models in  learning complicated signal distributions, modeling wireless channels, and  denoising and reconstructing distorted signals.   
First, fundamental working mechanism of diffusion models is introduced. Then the recent advances in applying diffusion models to wireless systems are reviewed. Next, two case studies are provided, where \emph{conditional diffusion models} (CDiff) are proposed for data reconstruction enhancement, covering both the conventional digital communication systems, as well as the semantic communication (SemCom) setups.     
The first case study highlights about 10 dB improvement in data reconstruction under low-SNR regimes, while mitigating the need to transmit redundant bits for error correction codes in digital systems. The second study further extends the case to a
SemCom setup, where \emph{diffusion autoencoders}  showcase  superior performance compared to legacy autoencoders and variational autoencoder (VAE) architectures. 
Finally,  future directions and existing challenges are discussed.  
\end{abstract}


\section{Introduction}\label{sec:Intro} 
\vspace{0mm}   
The swift progress in the realm of  artificial intelligence and machine learning  (AI/ML) has created  an  aspiration  to integrate AI into different functionalities of communication systems.     Nevertheless, traditional AI/ML models seem to fall short in delivering the anticipated performance. 
This stems from  the  complexities arising  from the massive volume of data from a multitude of devices, as well as the various use-cases.  
Generative AI (GenAI)  have been shown to be a powerful means to address this, exploiting their capabilities for learning intricate signal distributions, modeling and generation of complicated channels, and predictive modeling of complicated features \cite{GenAI_Telecom, GenAI_PHY}.

In this article, we focus on denoising  diffusion models and study their potentials for  wireless communication systems.  Distinguished from conventional generative models such as generative adversarial network (GANs) and variational autoencoders (VAEs), the idea behind diffusion models is to decompose data generation process over ``denoising'' steps, and gradually generate samples out of noise. The mechanism intuitively   highlights a strong relation to the overall goal of a communication system, i.e., to denoise and reconstruct information signals out of noisy ones.  
While the diffusion model research has  initially been focusing on visual AI, there has been an increased  interest in exploring its capabilities for communication systems   \cite{wcl, DM_WiFi, CGM_ChanEst, DM_for_E2EComm}.    
    
Diffusion models are utilized in \cite{DM_for_E2EComm} to augment channel datasets via synthetic channels, bypassing differentiable channel models for end-to-end (E2E) trainings of the communication system. The results highlight the performance  of diffusion models compared to {GANs} in both training and inference phases.    
Score-based diffusion models are proposed in \cite{CGM_ChanEst}  for  channel state information (CSI) estimation in multi-input-multi-output (MIMO) systems, utilizing their capabilities {to estimate the gradient of the log-prior} of high-dimensional wireless channels.  
The results highlight a competitive performance for out-of-distribution (OOD) scenarios compared to GANs and compressed sensing methods. 
A computation-efficient variant of diffusion models is proposed in \cite{wcl} for cell-free massive MIMO systems to  mitigate downlink interference, trading-off the computation complexity and communication accuracy.  Similar  diffusion model algorithm is employed in \cite{DM_WiFi}  to generate  fine-grained received signal strength (RSS) radio maps over WiFi, highlighting the performance compared to traditional models such as residual neural network architectures (ResNets). 

In this article,  we first provide the  fundamentals on how diffusion models work. We further review the recent advances, and future prospects in diffusion models being applied to different functionalities of the wireless communication pipeline. 
We highlight why and how diffusion models can address the problems in wireless domain.   
We further conduct  two case studies.  In the first study, we  employ a conditional denoising diffusion probabilistic model
(DDPM)  for wireless digital communication systems. Our study showcases how \emph{diffusion priors}  can replace the need for harsh channel codes, reducing the need for redundancy bits, thus improving the data rate as well as improving the reconstruction performance.   
As our second case study, we go beyond the traditional digital communication schemes, and study the applications of conditional diffusion models (CDiff) for semantic communication (SemCom) systems, where fully AI-based neural encoding and decoding modules handle the communication pipeline. We show how diffusion models  can be employed as a stand-alone neural decoder, replacing the  matched decoders conventionally used in legacy autoencoder-based architectures.

Even though there are prior works on outlining the applications of generative models for wireless communications \cite{GenAI_emerging} and \cite{GenAI_channel}, we keep our focus solely on diffusion models, as the most promising state-of-the-art in non-language generative models, and we neglect the redundant backgrounds on the widely-explored conventional generative models such as GANs or VAEs.  
We aim to advance the understanding of how diffusion models specifically fit into wireless systems, offering technical insights and design implications, as well as shaping new research directions.    
We show the unique features of  diffusion models which make them suitable for wireless applications. These features include robustness to noise (low-SNRs), ability to model complicated  distributions, and iterative refinement mechanism that is aligned with channel estimation or decoding tasks.  
Unlike prior articles such as \cite{GenAI_channel} which only focus on one specific aspect of diffusion models, 
we consider a \emph{holistic perspective,} investigating how diffusion models can be integrated across the \textbf{entire wireless data communication pipeline,} including for the AI transmitter, channel modeling, and AI receivers.  

In Section \ref{sec:DDPM} we first  introduce the fundamental concepts of diffusion models.   
Next, in Section \ref{sec:Applic}, we shed light on some of the potential roles that diffusion models can play in wireless systems.     
We then delve into studying the use-cases of diffusion models in wireless systems in Section \ref{sec:case_study}.     
We finally discuss the future directions  and open issues in Section \ref{sec:Challenges_open}, and conclude the article in Section \ref{sec:concl}.

\begin{figure}
	\vspace{0mm}
	\centering
	\includegraphics
	[width=3.45in,height=3.1in,trim={0.0in 0.0in 0.0in  0.0in},clip]{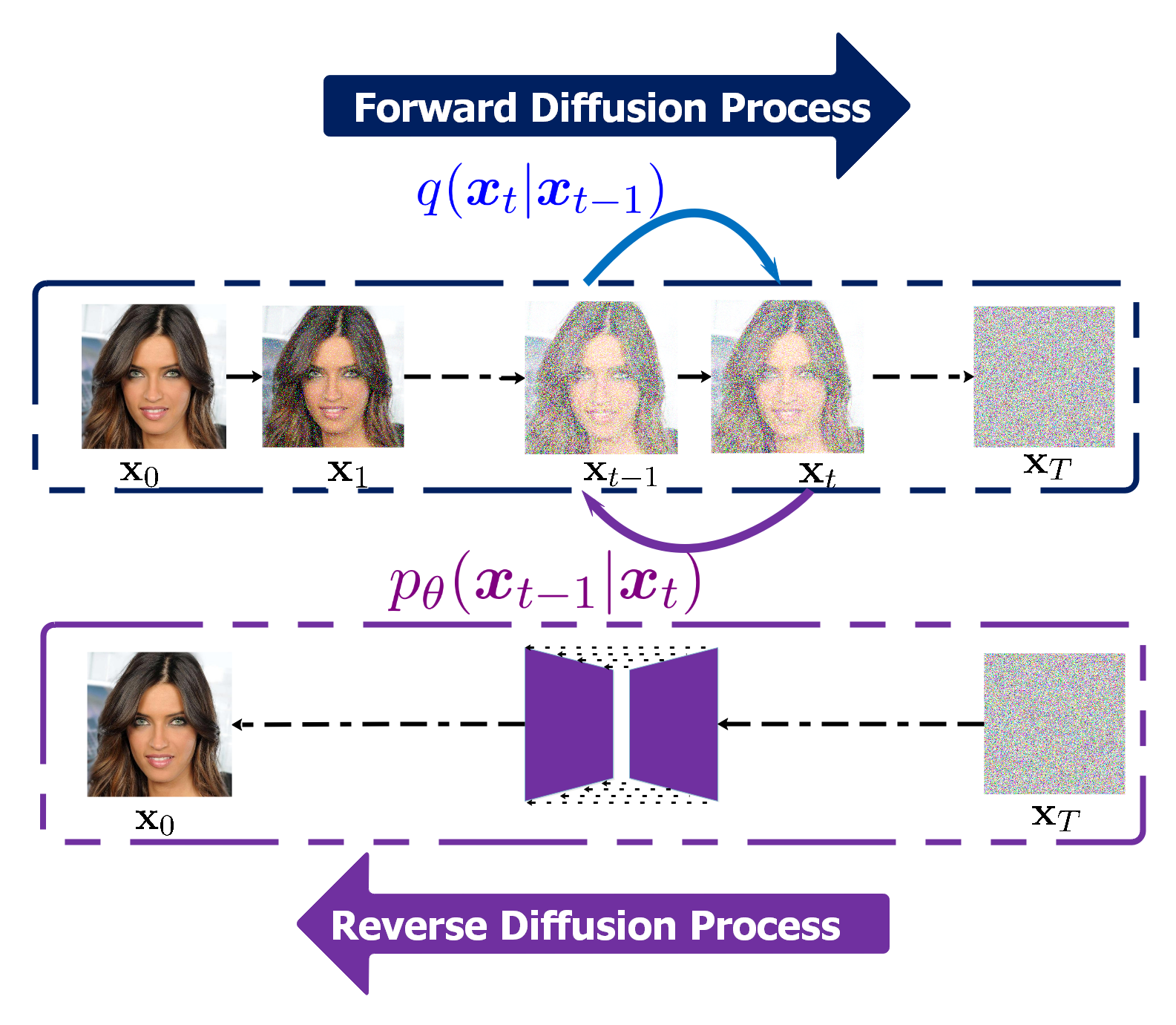}
	\caption{Working mechanism of denoising diffusion  models.  A Markov chain is defined to mimic the forward diffusion process, during which random noise is purposefully added to the original data. Then in a reverse process, a neural model learns to reconstruct the ground-truth out of noise. 
 }
	\label{fig:DM}
 \vspace{-2mm}
\end{figure}

\begin{figure*}[!tbph]
	\vspace{0mm}
	\centering
	\includegraphics
	[width=6.4in,height=3.45in,trim={0.0in 0.0in 0.0in  0.0in},clip]{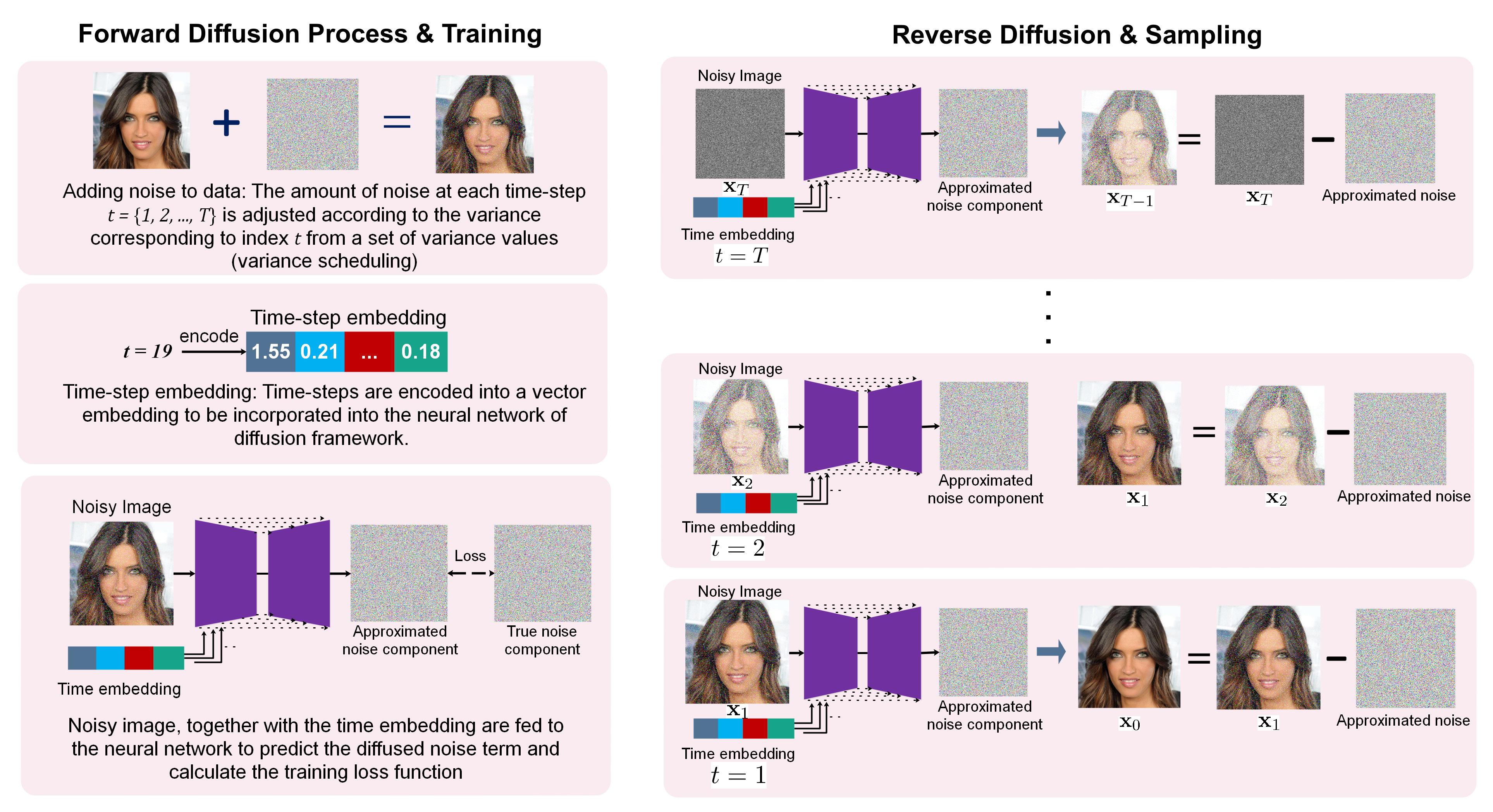}
	\vspace{2mm}\caption{{Details of forward diffusion process, training the diffusion model, and the reverse diffusion process.}}
	\label{fig:details}
 \vspace{-2mm}
\end{figure*}

\section{Fundamentals of Diffusion Models}\label{sec:DDPM} 
In this section, we provide the fundamentals and key concepts behind diffusion-based generative modeling.  
Diffusion models are a class of  state-of-the-art  probabilistic generative models that have showcased high sample quality, strong mode coverage, and sample diversity \cite{DM_Ho}.   
\emph{Like the mythical bird that rises from ashes, diffusion model is  characterized  by the transition from chaos (noise) to creation (data generation).}

The mechanism  behind  diffusion models  is to decompose  data  generation process over small ``denoising'' steps, through which the diffusion model   corrects itself and gradually generates desired samples.    
The key idea  is that \emph{if we could develop a machine learning  (ML) algorithm that is capable of learning the {systematic decay of information}  due to noise, 
then it should be possible to ``reverse'' the process and recover the information back from the noisy/erroneous data.}  
This is fundamentally different from  previous generative models---instead of trying to directly learn  distribution like in GANs, or learning a latent space embedding like in VAEs, diffusion models 
diffuse  data samples by adding noise using a Gaussian kernel,  and then  try to ``decode'' the information via ``denoising'' the perturbed data in a hierarchical fashion. 
In this article,   we are mainly focused    
on DDPMs, as one of the widely-adopted  state-of-the-arts  proposed by Ho \emph{et. al,} 2020 \cite{DM_Ho}.

As illustrated in Fig. \ref{fig:DM}, diffusion models are comprised of two processes, namely  forward diffusion process  and parametrized reverse process. 
Within the forward process, data is mapped into noise by  using a Gaussian diffusion kernel, gradually  perturbing the input data.  
In other words, at each step of the forward process,  Gaussian noise is incrementally added to the data.
The second process is a parametrized reverse process that aims to  ``undo'' the forward diffusion in an  iterative denoising mechanism (a.k.a sampling), regenerating the original  samples out of noisy input.  
Drawing  a  sample  
from an unknown (and possibly  complicated)    
distribution, the forward diffusion process 
is defined by adding  Gaussian noise 
with adjustable variance at each time-steps, which is known as  variance scheduling. Data samples  gradually lose their  distinguishable features as the time-step goes on, such that  after a sufficient  number of steps, they approach an isotropic Gaussian distribution \cite{DM_Ho}.  
The variance scheduling is designed beforehand, which  results in samples with different noise levels over time as illustrated  in Fig. \ref{fig:DM}.   
Intuitively, this implies  that by properly designing the variance scheduling in the diffusion process,  the model  would be able to ``see''  different structures of the distortion noise during  training, 
making it \emph{robust} against a wide range of distortions  when sampling.       
Then a neural network is trained, with the noisy samples along with an embedding of the time-steps as input, with the objective of estimating the noise vector in the distorted data.    
The diffusion model is trained based on a loss function that  calculates an error measure, e.g., mean-squared error (MSE), between the true diffused noise term and the predicted noise as shown in Fig. \ref{fig:details}. Finally, during the inference (sampling), the reverse diffusion process is run to   
regenerate  the true samples from  noisy signals.

\section{Recent Advances: Applications and Use-cases}\label{sec:Applic}   

\subsection{Beamforming and Downlink Transmission Policy} 
Diffusion models  can be used at AI transmitters to learn the optimal transmission policy and the corresponding beamforming vectors.  
In this scenario, the state and action for learning the optimal policy are the estimated wireless channels and the beamforming vector, respectively.  
The optimal beamforming requires first obtaining accurate  estimation of the wireless channel matrix   and then calculating the corresponding optimal beamforming matrix. 
A diffusion model can then be trained to learn the optimal policy distribution  \cite{DM_multiple},  taking the estimated channels as  the state input to the diffusion model, while 
adding noise on each diffusion step   
 can facilitate escaping the local optimal solutions and explore, in a fine-grained manner, the  better actions (beamforming vectors) at each iteration  until convergence.      
 A case study has been shown in \cite{DM_multiple} by learning digital beamforming
at the base station for transmitting to four single-antenna users. 
The channel coefficients from the base station to the users are assumed to
be accurately estimated, and the power allocation to the users is learned via diffusion models. Having the optimal power allocation policy, the beamforming matrix is derived following the signal processing-based beamforming formulas.

\subsection{Radio Map Prediction} 
Radio maps represent radio attributes 
of interest, such as received signal strength (RSS),  across a geographical region. 
Fine-grained, high-quality radio maps can provide network with predictive capabilities 
such as adaptively adjusting transmission parameters according to spatio-temporal radio map variations. 
Diffusion models can be leveraged at the transmitter (e.g., the gNB of a cellular system) to learn, estimate, and  generate high-fidelity radio maps. 
Taking the building maps (layouts) as prior information, and the 
collected UE radio measurements    
as  \emph{conditioning} input,  
diffusion models can  generate radio maps with high sample quality. 
Technically speaking, this can be viewed as solving an inverse problem \cite{inverse}, where the diffusion models learn the mapping from sparse radio measurements of UEs in the field to the full radio map.  
Diffusion-based radio map estimation  has been studied in \cite{DM_WiFi},
using RSS radio map dataset simulated in 5750 MHz (WiFi 5G) frequency band, and DDPMs as the diffusion model, where between 0.02\% to 0.31\% of the radio samples were collected to simulate sparse measurements.   
Ultimate technical effect would be the accurate prediction of the RSS distribution in the vicinity of the transmitters can increase the ``environmental-awareness'' of future AI-based transmitters.

\subsection{Synthetic Channel Generation for End-to-End Training}
Future AI-native systems are assumed to extensively employ machine learning algorithms that  jointly optimize  the communication blocks in an E2E fashion.  
Training such systems with gradient-based optimizers requires  the channel to be known and differentiable for ``backpropagating'' through the system.  However, communication channels do not necessarily follow a tractable and differentiable  mathematical models, and we might not have  access to the channel distributions in real-world scenarios, but to the samples from it.  In such scenarios, diffusion models are considered  as a promising solution to generate high-quality samples of the wireless channel. 
Diffusion models  can be employed  to synthetically  generate high-quality channel samples and/or their  derivatives, which is used for E2E gradient-based training. Such schemes can  consist of the neural  encoder, the  neural  decoder, and  the diffusion-based channel generator implemented in between the encoder and the decoder blocks, replacing the real channel.    
Then during the E2E training,  the diffusion model is trained using the encoder's output  (randomly sampled encoded channel inputs), and the corresponding channel outputs, and then the neural encoder-decoder pair is trained with the backpropagation path made by the channel realizations generated by the diffusion model. 
It is empirically  shown in \cite{DM_for_E2EComm} that diffusion models can be employed in  E2E wireless communications, outperforming  GANs with a more stable training procedure and better generalization performance.

\subsection{CSI Estimation and Overhead Reduction} 
Diffusion models can enable high-accuracy and low-overhead CSI estimation for MIMO communications. The role of diffusion models can be seen from different perspectives: I)  Diffusion models can effectively learn the channel score function, i.e., the gradient of the log-prior distribution of channels, given the received pilots. Such diffusion models are called  score-based models in the GenAI literature \cite{CGM_ChanEst}.  
Using  pretrained score-based models in conjunction with the received pilots, estimated CSIs can be can iteratively updated  in an unsupervised  manner, using the so-called  \emph{posterior sampling} method \cite{CGM_ChanEst};    
II) Diffusion-based CSI estimation can be seen as solving an ``inverse problem'' with the received de-modulation  reference signal (DMRS) treated as \emph{sparse measurements}. Transmitted DMRS pilots go through a typically non-invertible ``forward process,'' where the objective of CSI estimation is to ``invert'' the process and reconstruct the full CSI matrix from the observed sparse DMRS pilots.  
In this setup, diffusion models' fine-grained and high-quality sampling is beneficial, allowing to relax the requirements on fixed DMRS pattern across antennas, while  fine-grained  CSI recovery is still possible with sparser (and perhaps dynamic) DMRS patterns.  This can obviously reduce the CSI estimation overhead and saves physical layer radio resources;  III) In addition, diffusion models are inherently robust against noise.  For CSI estimation use-case, this leads to robustness against \emph{pilot contamination}. That is,  diffusion models can recover accurate channels
from noisy pilots, even when affected by interference and/or pilot contamination.  IV) Finally,  diffusion models by nature maintain a relatively simple conditioning mechanism, through which varying wireless conditions can be incorporated into the  model to steer the generation (sampling) process. To exploit this capability for CSI estimation, diffusion-based channel estimators  can be pre-trained on simulated data such as the available 3GPP channel  models or  from ray-tracing simulators, and then fine-tuned in real-world environments via conditioning them on varying channel/environment attributes.

\subsection{Data  Reconstruction Enhancement}
The idea behind diffusion models, as elaborated earlier, is to break down the data generation process into a sequence of ``denoising'' steps, gradually producing samples from noise. This characteristic naturally aligns with the fundamental goal of a receiver, i.e.,  to recover and reconstruct information signals from noisy observations. Accordingly, diffusion models as powerful denoisers, can be used to
clean noisy signals received over communication channels   leading to improved signal
quality \cite{CDiff_TMLCN}.  
Data transmission can be seen as a  forward diffusion  process \cite{wcl}, where noisy encodings of the source data are received, and the receiver  uses the diffusion sampling to denoise, decode, and regenerate  the
original data, optimizing the overall communication pipeline for robustness and efficiency.  This is particularly pronounced in extreme channel conditions with poor connectivity due to low signal-to-noise ratio (SNR) or highly mismatched transmitter and receiver due to residual RF impairments, where even error correction capabilities might not be satisfactory.  Diffusion models at the receiver of a communication system can also help learn the multi-user interference signals, predict, and remove them.  
In  \cite{CDiff_TMLCN}, it has been shown that the \emph{generative prior} of diffusion models can be exploited to enhance data reconstruction at the receiver.  More details are followed in the next section where we provide two case studies on the use of diffusion models at the receiver.

\subsection{Channel Code Design}
Diffusion models can also be used to directly design 
channel decoding schemes. 
Channel codeword corruption can be viewed as a series of forward diffusion process to be reversed via diffusion model. 
DDPMs are likelihood-based generative models, hence, they seem to have close relation to the principals of maximum likelihood decoding, while their inherent iterative scheme mimics that of legacy decoders.   
In  \cite{ECC}, a DDPM-based error correction algorithm is designed for the soft decoding of linear codes at arbitrary block lengths,  
where the diffusion decoding is conditioned on the number of parity errors to incorporate the level of corruption into the denoising and decoding process at each given step. Further studies in this direction could be envisioned at the intersection of diffusion model research and coding theory.


\begin{figure} 
	\vspace{0mm}
	\centering
	\includegraphics
	[width=3.45in,height= 2.55 in,trim={0.0in 0.0in 0.0in  0.0in},clip]{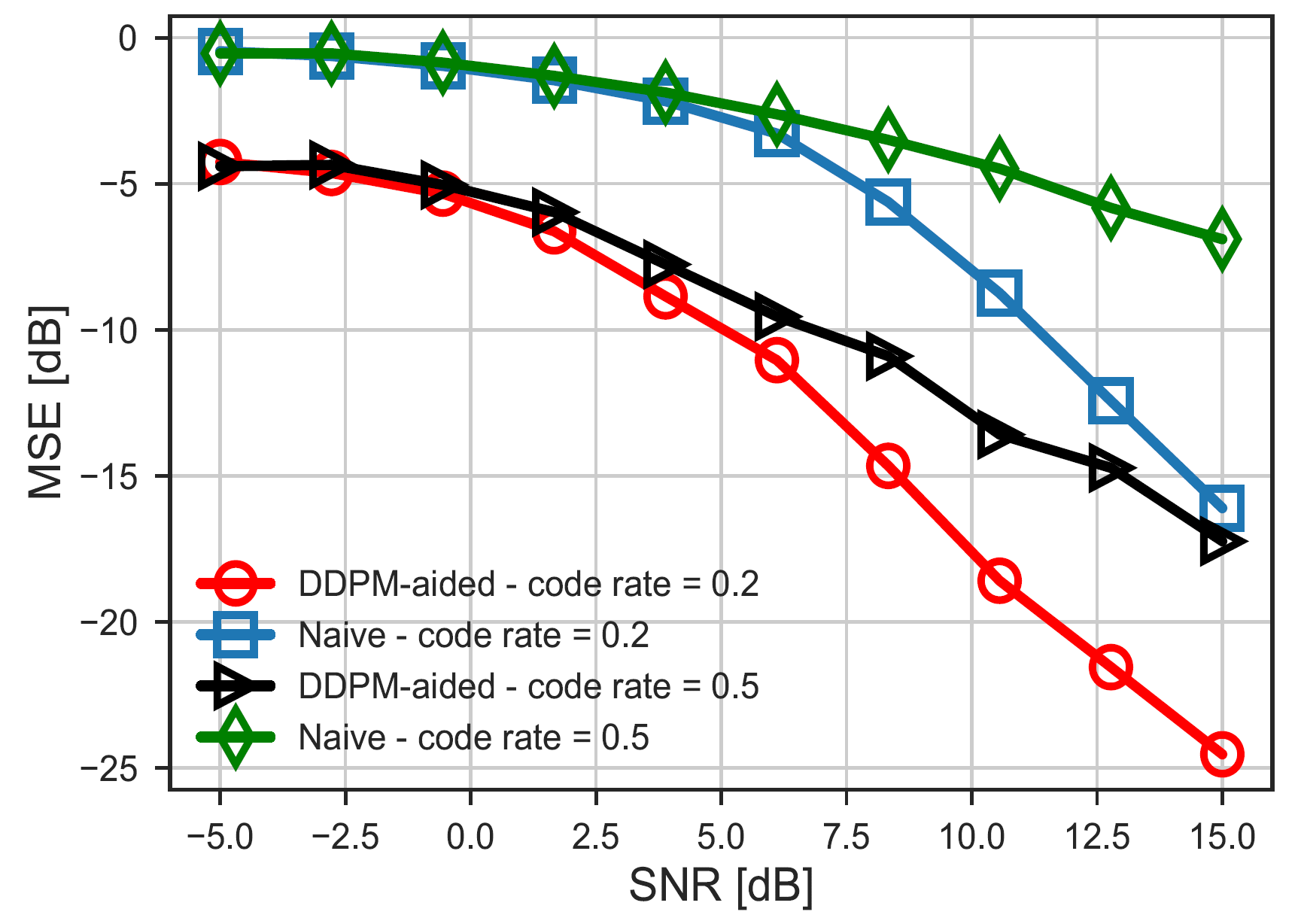}
	\caption{Case study 1: Reconstruction performance of DDPM-aided receiver \cite{CDiff_TMLCN}. Enhanced reconstruction performance and lower channel coding rate is achieved compared to the conventional digital communication receiver.}
	\label{fig:Use_case1}
 \vspace{-2mm}
\end{figure}

\section{Case Studies}\label{sec:case_study} 

\subsection{Diffusion Model-Aided Receiver for Digital Communication Systems} 
In this case study,  
we propose conditional DDPMs for enhancing the receiver's reconstruction performance in digital communication schemes.  
The key idea is that, having a generative model at the receiver, not all the packets are required to be received error-free. Rather, \emph{generative priors} can compensate for the erroneous/lost packets, regenerating them locally at the receiver.  
Thus, instead of employing harsh channel codes that reduce the information rate, transmitter can send the datastreams with less complicated  correction codes and lower code rates.    \emph{Diffusion priors} are then  exploited locally at the receiver to enhance the reconstruction.  
This is particularly pronounced in
extreme channel conditions with poor connectivity due to low SNR or residual RF impairments \cite{CDiff_TMLCN}. In such scenarios, error correction  codes might not be
able to correct the mismatches, or complicated codes might be needed at the cost of decreasing the information rate.

The proposed DDPM-assisted  receiver is employed  in conjunction with the NVIDIA  Sionna simulator\footnote{\url{https://github.com/NVlabs/sionna}} {by plugging the trained diffusion model to  the end of the receiver chain, i.e., after the signal detection and channel decoding.}   We consider the image transmission use-case using MNIST dataset, where   
the pixel values are mapped to binary bitstreams, encoded via low density parity check coding (LDPC), and mapped to 64-QAM modulation symbols to be sent over wireless channel. After decoding the bits and converting the reconstructed bits to  reconstructed image, a DDPM is deployed for conditional denoising and image reconstruction enhancement.  
{For the choice of neural network, we follow the architecture proposed in \cite{DM_Ho}.   
While this architecture assumes that the denoising process starts with
isotropic Gaussian noise, in a practical wireless
system the received/decoded signals are a degraded version
of the ground-truth information signals, not necessarily a pure
isotropic Gaussian noise. 
Moreover, the degradation might be different
in different distortion conditions such as different SNRs.  To address this, 
the noisy/erroneous  received signals are incorporated as the ``conditioning'' into the denoising framework, and the model  reconstructs the ground-truths from the noisy versions instead of an isotropic Gaussian noise. 
Training was carried out for 10 epochs over 200 steps, using  adaptive moment estimation (Adam) optimizer.

Fig. \ref{fig:Use_case1}   demonstrates the   reconstruction performance.  
The figure shows that under extreme conditions  (low-SNR and mid-SNR regimes), a naive scheme (without any  diffusion prior) cannot perform well in terms of the image reconstruction, even if the code rate is decreased from $0.5$ to $0.2$.  When the conditional DDPM is employed, the reconstruction is enhanced significantly, by more than 10 dB.  
The figure suggests that instead of employing error correction codes with lower code rates,  our proposed scheme can be employed while preserving the data rate. As can be seen from the figure, more than $5$ dB performance improvement, as well as $2.5$ times higher data rates can be achieved under low SNR regimes with our proposed scheme.


\begin{figure} 
	\vspace{0mm}
	\centering
	\includegraphics
	[width=3.2in,height=2.1in,trim={0.0in 0.0in 0.0in  0.0in},clip]{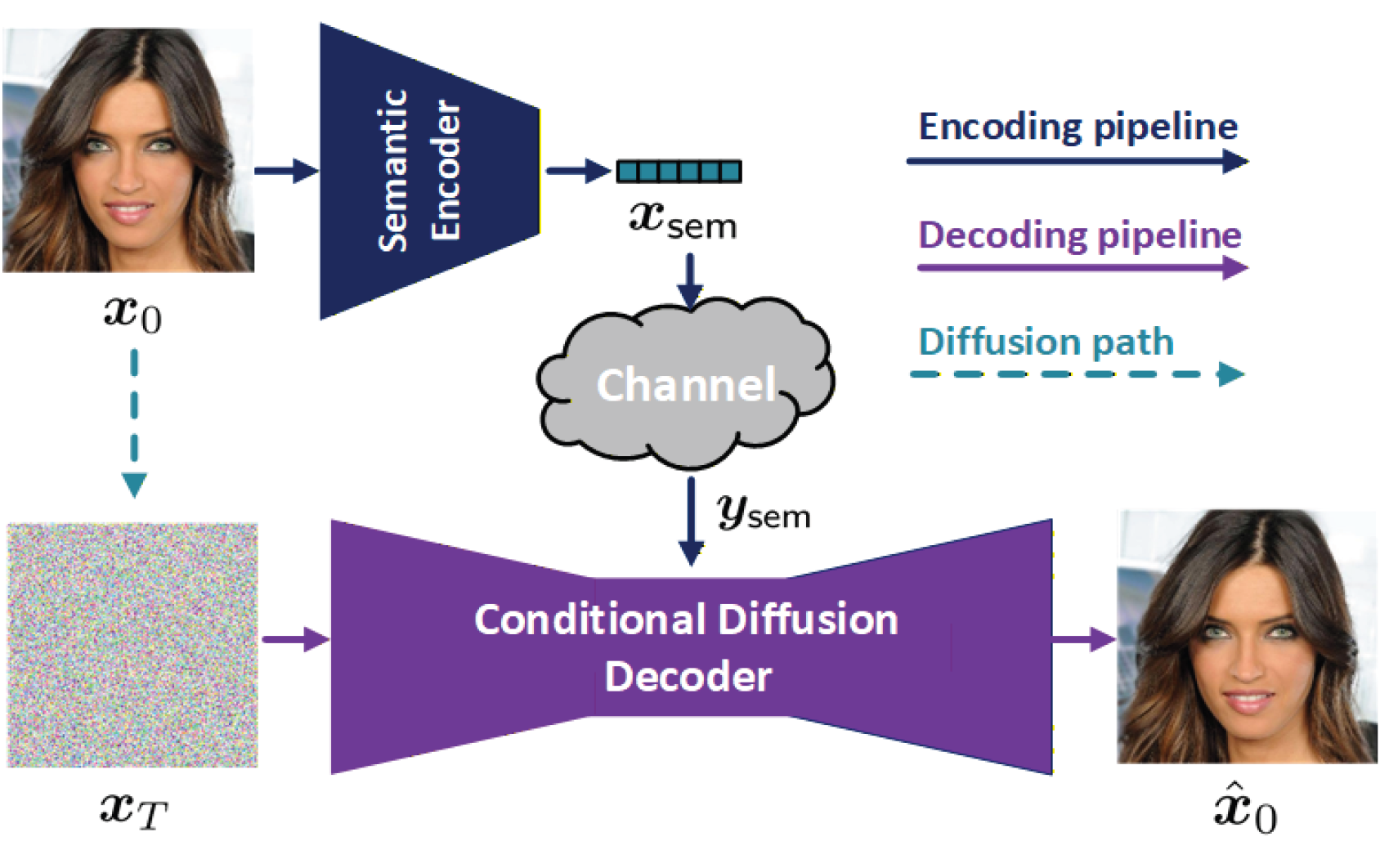}
	\caption{System model of the SemCom case study.}  
	\label{fig:Use_case2_sysMod}
 \vspace{-2mm}
\end{figure}

\begin{figure} 
	\vspace{0mm}
	\centering
	\includegraphics
	[width=3.45in,height=2.55in,trim={0.0in 0.0in 0.0in  0.0in},clip]{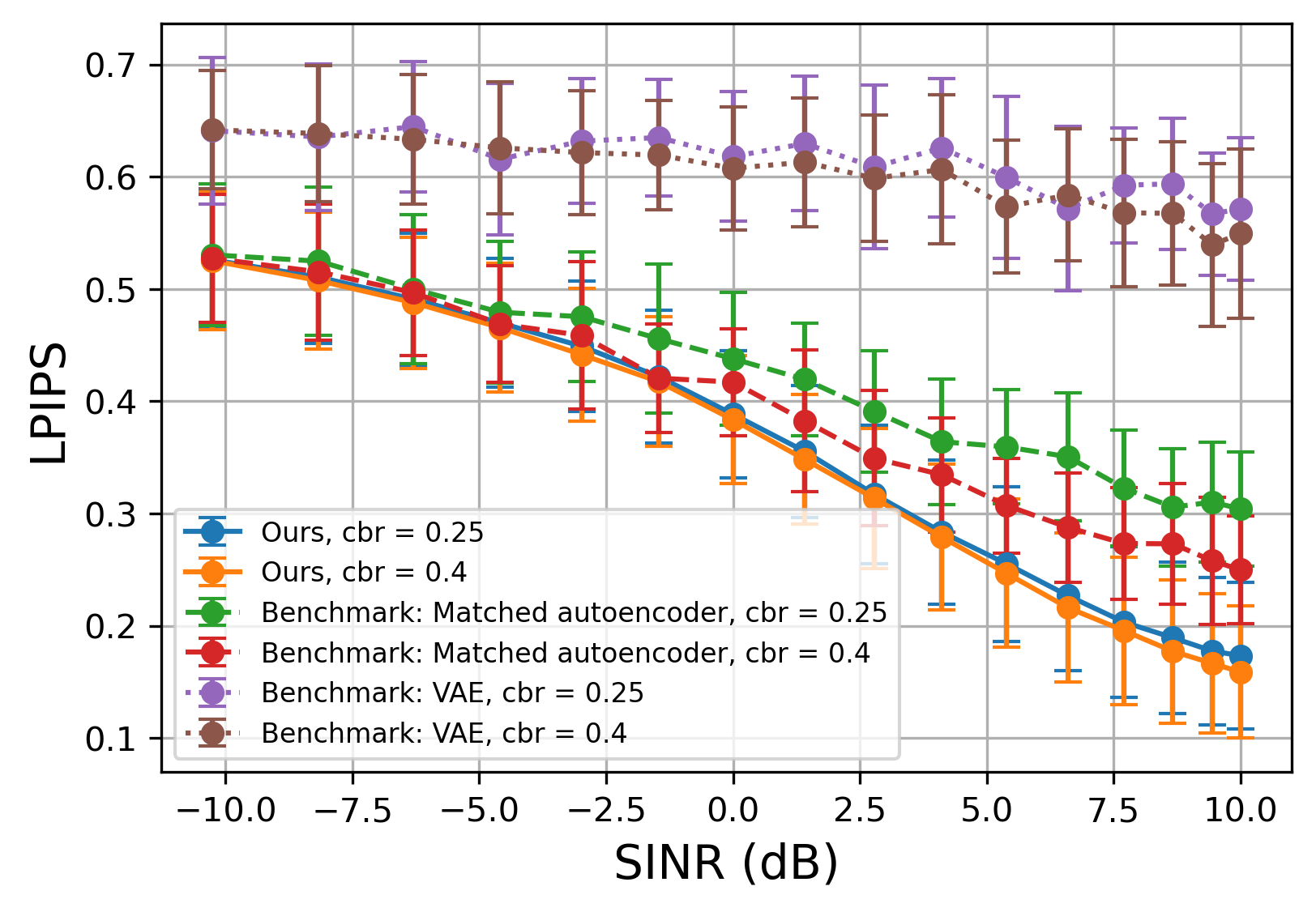}
	\caption{Case study 2: Multi-user SemCom performance of the proposed diffusion autoencoder for different CBRs \cite{NeurIPS}.  
   Enhancements can be achieved  compared to the legacy autoencoders and VAE benchmarks.}  
	\label{fig:Use_case2}
 \vspace{-2mm}
\end{figure}

\subsection{Conditional Diffusion Autoencoders for Semantic Communication Systems} 

We further extend the case studies to a fully neural-network based communication under the SemCom framework.  While in the previous case study, diffusion model learned a ``noisy-to-clean'' mapping,   
here the idea is to learn  a ``semantic-to-clean'' mapping, from the semantic space to the ground-truth probability distribution.  
A semantic neural encoder extracts the source semantics, and sends the semantic latents over the wireless channel, according to a channel bandwidth ratio (CBR) $k/n$, which is defined as the ratio between the source data dimension $n$ (a.k.a {source bandwidth}), and the channel dimension $k$  (a.k.a {channel bandwidth}). 
At the decoder side, a conditional DDPM is proposed for the neural decoder, exploiting the source distribution for signal-space denoising, while the received noisy semantics are incorporated as the conditioning input to the diffusion decoder.  This way, semantic latents are distilled into the decoding process, which helps ``steer''
the decoding process towards the semantics intended by the transmitter \cite{NeurIPS}.  Fig. \ref{fig:Use_case2_sysMod} illustrates the system model.   
It is worth mentioning that 
SemCom systems traditionally relied on autoencoder architectures, where the encoding functionality is tightly coupled to a matched decoder. 
This causes serious \emph{scalability issues} in practice. With our proposed architecture, this is handled smartly.  

Our neural encoder consists of four convolutional layers 
with kernel size $3 \times 3$, stride $2$ and padding $1$, 
each  
followed by batch normalization and a ReLU activation. 
Downsamples input is then projected through a fully connected layer to produce a latent vector of length $2k$, 
(with real and imaginary parts stored separately). 
The latent vector is further normalized to have unit $\ell_2$-norm to satisfy the average power constraint.  
The implementation of our diffusion decoder model is inherited from  the seminal DDPM framework  \cite{DM_Ho}
converted to PyTorch. 
We further incorporate the conditioning mechanism into this framework, where the conditioning mechanism 
is developed  upon CDiffuSE architecture\footnote{\url{https://github.com/neillu23/CDiffuSE}}, and adapted to the image generation task.


Fig. \ref{fig:Use_case2} demonstrates the performance of the proposed scheme in terms of learned perceptual image patch similarity (LPIPS) metric over CIFAR-10 dataset, where lower  value indicates higher perceptual similarity between the ground-truths and the decoded version.     
We study the scenario of multi-user SemCom, where the  decoder receives a
superimposed version of semantic vectors from
two neural encoders, each independently transmitting their
own i.i.d data over the same
channel. 
The figure highlights the outperformance of the diffusion autoencoder compared to VAEs and also legacy autoencoders with matched encoding-decoding architectures. 
The benchmarks maintain the same  encoder  architecture  as  the one employed for our scheme, and the decoder counterpart mirrors the encoding functionality for the autoencoder benchmark (matched decoder).  For the VAE benchmark,  features are further passed through two parallel linear layers which respectively estimate the mean   and log-variance of the posterior distribution over latent variables.
Notably,  VAEs do not maintain any ``fine-grained'' denoising mechanism to explore over probability distributions. Moreover, casting a standard normal distribution on the prior distribution of the latent space further compromises the decodability quality of semantics. On the other hand, our proposed diffusion autoencoder does not impose any constraint, allowing the neural encoder to obtain the semantics as expressive as it can.  
Our proposed scheme will have huge merits compared to the traditional autoencoder-based schemes. 
With our diffusion-based decoder model at the receiver (deployed e.g.,  at the gNB of a cellular system), the requirement to match the neural encoder and decoder pairs can be dropped.  This is because the built-in conditioning mechanism of diffusion models is able to flexibly adapt to variable-length latent embeddings  corresponding to different semantic latents. 
Therefore, there is no need to match the decoder architecture to every single neural encoder for each transmitter (e.g., the terminal devices in a cellular system).

\section{Future Directions  \&  Open Issues}\label{sec:Challenges_open}    
\subsubsection{Joint Sensing \& Communications (JSAC)}   
Diffusion models can  be utilized to  optimize AI-based  JSAC receivers.   
Diffusion models are capable of effectively learning and separating superimposed signals coming from multiple 
sources \cite{mit}.  
This can be seen as learning the posterior distribution of  the communication components as well as the  sensing components using pre-trained diffusion models.  
Such a pre-trained diffusion model can contain separately-trained statistical priors of communication and sensing  source signals, which will then be used to approximate the corresponding score functions used in the maximum a posteriori (MAP) detection.   
In practice,  time-series diffusion models 
(such as DiffWave architecture\footnote{\url{https://github.com/lmnt-com/diffwave}})  can be  trained in  waveform domain to learn the {generative priors} of source signals in the form of score functions. Such pre-trained score functions can then be used in each iteration of the MAP estimation for communication/sensing signal detection and recovery.    

\subsubsection{Network Digital Twins} 
Moving beyond radio maps discussed in \ref{sec:case_study},  diffusion models can be a promising solution for processing and generation of massive amount of high-quality synthetic data as the replicas of the real-world. For instance, a gNB of a cellular system may employ diffusion models, which take different data modalities such as 2D/3D images of the environment, sensory measurements such as locations, RSSs, and/or 3D point-clouds as input, 
and accordingly generate high-fidelity digital replica of the radio environment for the network.   
One important technical advantage would be the reduction in the amount of information needed to be transferred from the multi-sensory  devices to the gNB for building  such a digital twin.
This is thanks to the ability of diffusion models in coming up with a high-fidelity ``picture'' out of corrupted/missing data, implying that  sparse measurements of distributed UEs might help in generating a high-quality digital twin of the environment.  
Moreover, diffusion models are ``condition-friendly,'' i.e.,  they can adapt, and react fast to environmental condition changes. 
Hence, when a new radio condition  happens, it can be flexibly ``embedded'' 
into the model to ``guide'' the sampling process towards the direction matching the conditions.

\subsubsection{Distributed Diffusion Models and Privacy}  
Diffusion models typically maintain multiple iterations of denoising. This potentially large number of sampling steps results in generating fine-grained high-quality samples, however, 
it also comes with computation complexity and energy consumption.  
Distributing the sampling process over multiple processing nodes using parallel computing techniques can significantly  improve the computation efficiency. In this regard, various design principles should be studied, including the signaling coordination among the computing nodes (distributed  UEs) and the aggregator node (e.g., the gNB in a cellular system), scheduling and clustering of the computing nodes. e.g.,  based on the saptio-temporal correlations among a set of co-located UEs, as well as the energy efficiency optimizations with respect to radio resources.  
Looking to this problem from another angle, each data sample within the intermediate denoising steps contains some level of noise. This may provide an inherent level of privacy preservation against intermediate computing nodes. Therefore, there might be some fundamental relations between diffusion sampling and  \emph{differential privacy,} which requires  more extensive studies in the future.

\subsubsection{Theoretical Aspects and Formal Methods}  
Diffusion models, like  many other generative AI algorithms,  are  \emph{probabilistic models}, thus {non-deterministic}  by nature.  Consequently, the same input might  yield different outputs across multiple sampling runs. In-depth studies seem to be needed with the aim of establishing formal guarantees on output consistency or reliability.  
As we showed in \cite{NeurIPS}, it requires a deep understanding of diffusion model fundamentals,  combined with  probability theory, learning theory, and data science. 

\subsubsection*{Existing Challenges} 
Diffusion models suffer from a drawback with respect to  the long run-time for training and sampling. This might make it   challenging  to develop diffusion-based algorithms for the wireless communication systems with stringent  latency requirements.  For example, one important direction would be to study novel techniques to speed up the generation process of diffusion models  for mission-critical communication scenarios. 
Notably,  there is an active line of research on efficient AI, exploring  ways to accelerate the sampling process to come up with computation-efficient diffusion models.   
This problem  can also be viewed from another angle. As presented in our case studies, diffusion model-aided schemes can make the communication systems  radio resource-efficient. This is achieved by saving the physical resource blocks (and bandwidth), due to  sending only the minimal relevant information rather than the entire datastream in case of SemCom, or less redundancy bits in case of digital communication systems.   In other words, the processings are moved from the wireless physical layer to the ``AI layer,'' where the AI processing happens at the cloud, over-the-top (OTT) server, or other ``core network'' entities with richer compute resources.  
This actually makes the wireless systems more radio efficient, particularly beneficial for resource-constrained wireless systems.   
For  larger-scale developments of diffusion models,  other challenges might pop up, related to model update  and life-cycle management for employing large diffusion models, which should  also be considered.

\section{Conclusions}\label{sec:concl} 
\vspace{0mm}
In this paper, applications of  diffusion models for wireless systems have been studied.  
Fundamentals of diffusion models have been introduced, and the recent advances in applying them to wireless systems have been reviewed in details. Two case studies have been provided, leveraging conditional DDPMs,  where significant improvements have been observed under both the conventional digital communication systems, as well as the SemCom setups.       
Insights on future directions and existing  limitations towards the widespread adoption of diffusion models have also been discussed.

\end{document}